\documentclass[5p, final,times]{elsarticle}
\usepackage{lineno,hyperref}
\usepackage{xcolor}
\journal{arxiv.org}

\begin{document}
\begin{frontmatter}

\title{A novel active veto prototype detector with an inner target for improved rare event searches}

\author[mysecondaryaddress1,mysecondaryaddress2]{M. Chaudhuri}
\author[mymainaddress]{A. Jastram}
\author[mymainaddress]{G. Agnolet}
\author[mysecondaryaddress1,mysecondaryaddress2]{S. Banik}
\author[mymainaddress]{H. Chen}
\author[mysecondaryaddress1,mysecondaryaddress2]{V. Iyer}
\author[mysecondaryaddress1,mysecondaryaddress2]{V. K. S. Kashyap}
\author[mythirdaddress]{A. Kubik}
\author[mymainaddress]{M. Lee}
\author[mymainaddress]{R. Mahapatra}
\author[mymainaddress]{S. Maludze}
\author[mymainaddress]{N. Mirabolfathi}
\author[mymainaddress]{N. Mishra}
\author[mysecondaryaddress1,mysecondaryaddress2]{B. Mohanty \corref{mycorrespondingauthor}}
\cortext[mycorrespondingauthor]{Corresponding author}
\ead{bedanga@niser.ac.in}
\author[mymainaddress]{H. Neog}
\author[mymainaddress]{M. Platt}

\address[mysecondaryaddress1]{School of Physical Sciences, National Institute of Science Education and Research, Jatni 752050, India}
\address[mysecondaryaddress2]{Homi Bhabha National Institute, Training School Complex, Anushaktinagar, Mumbai 400094, India}
\address[mymainaddress]{Department of Physics \& Astronomy, Texas A\&M University, College Station, TX 77843, USA}

\address[mythirdaddress]{SNOLAB, Creighton Mine \#9, 1039 Regional Road 24, Sudbury, ON P3Y 1N2, Canada}

\begin{abstract}
 
  We report the fabrication and performance of an annular, cryogenic, phonon-mediated veto detector that can host an inner target detector, allowing substantial reduction in radiogenic backgrounds for rare event search experiments. A germanium veto detector of mass $\sim$500 g with an outer diameter of 76 mm and an inner diameter of 28 mm was produced. A 25 mm diameter germanium inner target detector of mass $\sim$10 g  was mounted inside the veto detector. The detector was designed using inputs from a GEANT4 based simulation, where it was modeled to be sandwiched between two germanium detectors. The simulation showed that the background rates (dominantly gamma interactions) could be reduced by $>$ 90$\%$, and that such an arrangement is sufficient for aggressive background reduction needed for neutrino and dark matter search experiments. During testing at the experimental site the veto detector prototype achieved a baseline resolution of 1.24 $\pm$ 0.02 keV while hosting a functional inner target detector. The baseline resolution of the inner target detector was 147 $\pm$ 2 eV. The detectors were operated at mK temperatures. The experimental results of an identical detector arrangement are in excellent agreement with the simulation.  
\end{abstract}

\end{frontmatter}

\section{Introduction}
\label{intro}
The study of backgrounds, its reduction and understanding is of great importance in rare event search experiments like dark matter search \cite{DMoverview}, Coherent Elastic Neutrino-Nucleus Scattering (CE{$\nu$}NS) \cite{CENSFreedman1, CENS2}, and Neutrinoless Double Beta Decay (NDBD) \cite{NDBD}. All these experiments need to tackle a very high background rate to identify weak signals.There are mainly two types of background sources present in these experiments: radiogenic and cosmogenic. To reduce the effect of cosmogenic backgrounds most of these experiments are conducted in underground laboratories. Radiogenic backgrounds originate from radioisotopes present in the material surrounding the detectors. They are dominantly gammas and neutrons. They must be shielded to an excellent level in order to achieve competitive science goals. The primary background reduction method in these experiments is shielding. There are two types of shielding, passive and active shielding. In passive shielding, aside from operating deep underground, a multi-layer hermetic shield is used to substantially reduce the rate of background gammas and neutrons. While this method greatly reduces background event rates, further reduction will help to improve the sensitivity of the experiment. One way to aggressively reduce the impact of backgrounds is by applying active shielding. In this paper, an active veto detector is instrumented in the immediate vicinity of the target detector as an active shielding and can tag and reduce coincident background events by $>$ 90\% that could pollute the experimental signal region.

\begin{figure*}[t]
\centering
\begin{minipage}[b]{0.45\linewidth}
\centering
\includegraphics[width=\linewidth]{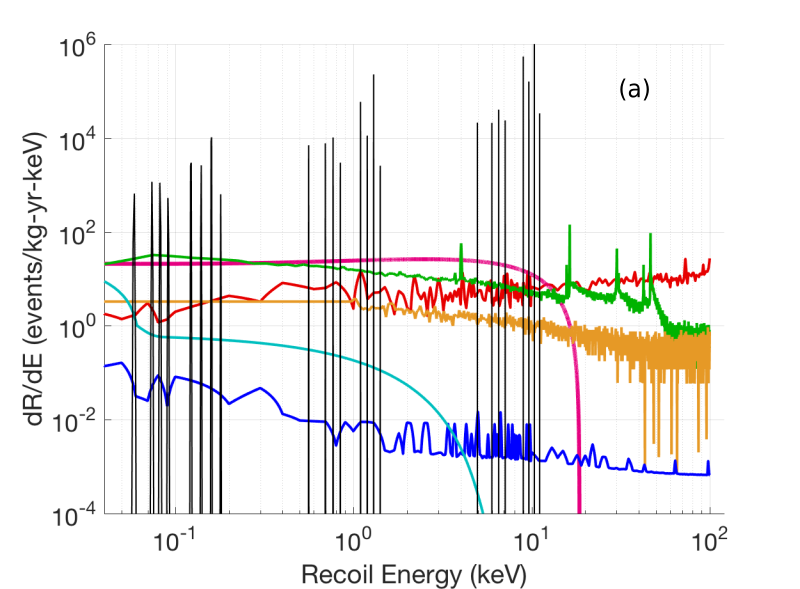}\\
\end{minipage}
\begin{minipage}[b]{0.45\linewidth}
\centering
\includegraphics[width=\linewidth]{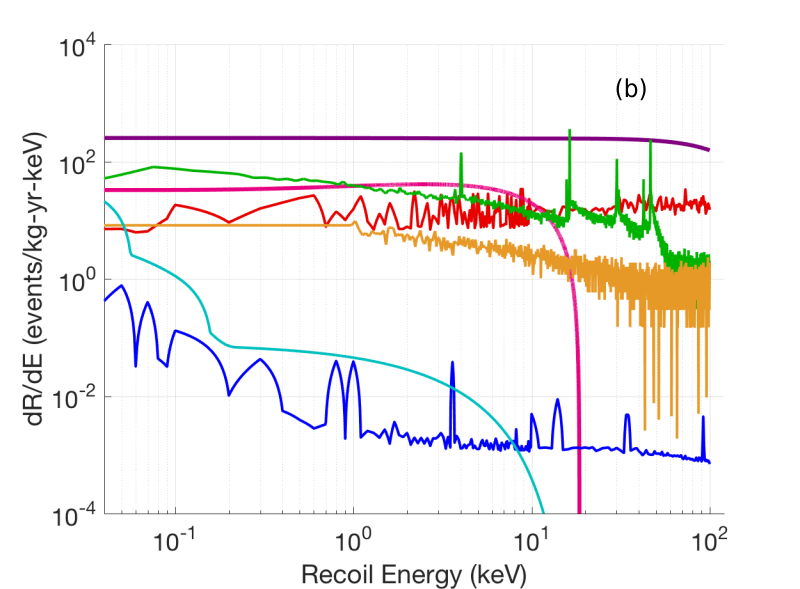}\\
\end{minipage}
\caption{Monte Carlo simulation of expected background rate in units of events/kg$\cdot$yr$\cdot$keV is shown as a function of recoil energy in the SuperCDMS SNOLAB experiment. (a) and (b) shows background rate in germanium and silicon detector respectively. $^{3}$H (pink) in germanium and $^{32}$Si (purple) in silicon are the main contributors to the backgrounds. The remaining components are Compton scatters from gamma rays (red), surface betas (green), decay daughters of $^{222}$Rn (orange), neutrons (blue) and coherent elastic neutrino-nucleus scattering (cyan). The black lines represent activation lines in germanium due to the electron capture process \cite{SNOLAB}.}
\label{fig:background}
\end{figure*}

\begin{figure}[h]
\centering
\includegraphics[width=0.9\linewidth]{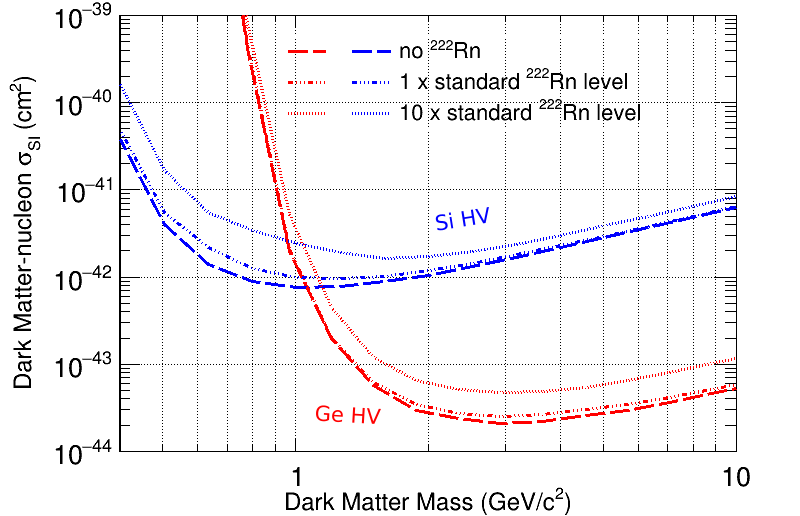}
\caption{Projected sensitivity for the SuperCDMS SNOLAB direct detection dark matter experiment showing variation in the limits due to different levels of \textsuperscript{222}Rn contamination \cite{SensitivityPlotter}. The vertical axis is the spin-independent dark matter-nucleon cross-section and the horizontal axis is the dark matter mass. The blue and red lines correspond to the \textsuperscript{222}Rn contamination in silicon and germanium HV detector respectively, showing the variation for zero (dashed line), standard value (dash dotted line) and 10 times the standard value (dotted line) of the \textsuperscript{222}Rn backgrounds where the standard value is taken as 50 nBq/cm$^{2}$ \cite{SNOLAB}.}
\label{fig:background2}
\end{figure}

For instance, in the Super Cryogenic Dark Matter Search (SuperCDMS) Soudan experiment, WIMP search sensitivity was limited by the decays of \textsuperscript{210}Pb \cite{Pb210} on the detector surface. This long-lived \textsuperscript{210}Pb (half-lives 22.3 years) is a decay daughter of \textsuperscript{222}Rn usually present in the surrounding environment and housing material like Cu. Three concerning radioisotopes are: \textsuperscript{210}Pb, \textsuperscript{210}Bi and \textsuperscript{210}Po which yield low-energy betas and X-rays ($\sim$46.5 keV). Decay of \textsuperscript{210}Po creates \textsuperscript{206}Pb which results in a nuclear recoil and can be observed in the signal region. In the upcoming SuperCDMS SNOLAB experiment, three of the primary backgrounds that may limit the sensitivity are (i) Compton scatters from the gamma rays, (ii) surface betas from surrounding materials, and (iii) recoils from decay radiation of \textsuperscript{210}Pb and it's progeny existing on the detector surfaces from radon exposure (see Fig. \ref{fig:background}) \cite{SNOLAB}. The \textsuperscript{210}Pb background has been studied extensively \cite{street2020removal}, and the improvements in the scientific reach have been simulated based on varying levels of its prevalence as shown in Fig. \ref{fig:background2}.

The design of the active veto detector is such that it would tag all the events as backgrounds which will deposit energy both in the active veto and inner target detector. It would thus reject essentially all of the recoils from \textsuperscript{210}Pb and surface beta events, while also reducing the Compton backgrounds expected in SuperCDMS SNOLAB by an order of magnitude. Reduction in backgrounds in this way will prove critical in making cutting edge scientific measurements in this field. Although, SuperCDMS detectors \cite{SupercdmsSudan} were the initial motivation for this veto detector design, this could also be benificial for other rare events search experiments with similar detector technology.

\section{Simulation}
\label{simulation}

\begin{figure*}[t]
  \centering
  \begin{minipage}[b]{0.38\linewidth}
    \centering  
    \includegraphics[width=\linewidth]{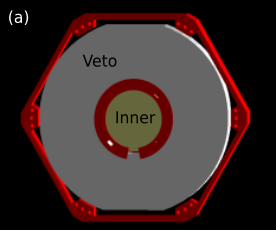}\\
  \end{minipage}\hfill
  \begin{minipage}[b]{0.315\linewidth}
    \centering
    \includegraphics[width=\linewidth]{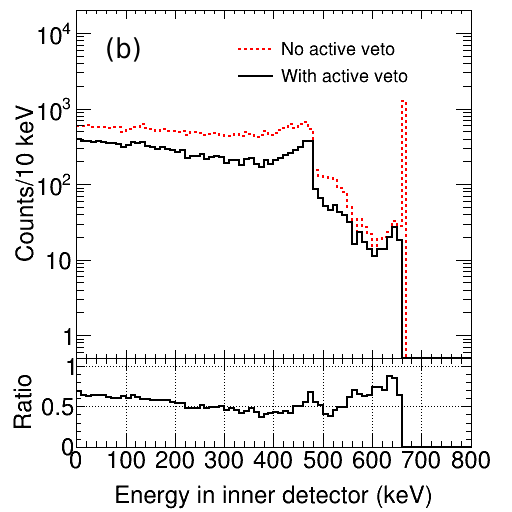}\\
  \end{minipage}
  \begin{minipage}[b]{0.28\linewidth}
    \centering
    \includegraphics[width=\linewidth]{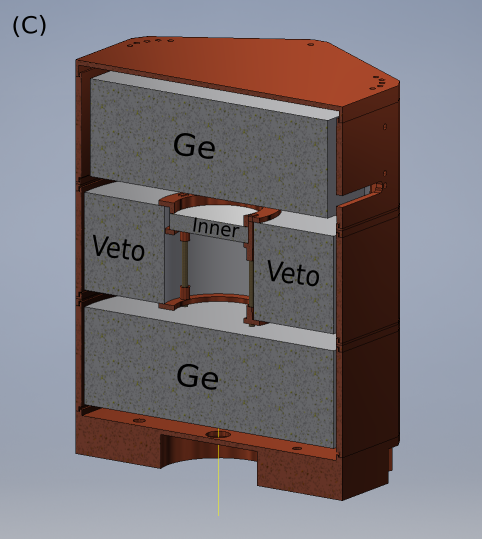}\\
  \end{minipage}

  \caption{ (a) A picture of an active veto detector with an inner target detector used in the simulation. The thickness of the veto detector is 25 mm, and its outer and inner diameters are 76\,mm and 28\,mm, respectively. Whereas the thickness and the diameter of the inner detector are 4 mm and 25 mm, respectively. (b) Simulated energy deposition spectrum (upper panel) for a $^{137}$Cs source and tagging efficiency (lower panel) as a function of energy deposition in the inner detector. The red dashed line in the figure shows the energy deposition spectrum in the inner detector before tagging and the solid black line after tagging with the veto. It shows that tagging efficiency by the veto detector is expected to be between 50 - 80\%. (c) A cross-sectional view of the detector stack used in simulation and drawn in CAD. The experimental result with a similar detector arrangement is discussed in Section \ref{results}.}
  \label{fig:Cs_simu}
\end{figure*}

\begin{figure*}[h]
\centering
\begin{minipage}[b]{0.45\linewidth}
\centering
\includegraphics[width=\linewidth]{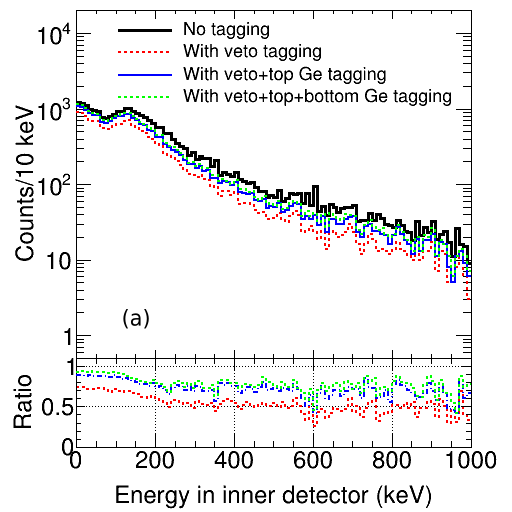}\\
\end{minipage}
\begin{minipage}[b]{0.45\linewidth}
\centering
\includegraphics[width=\linewidth]{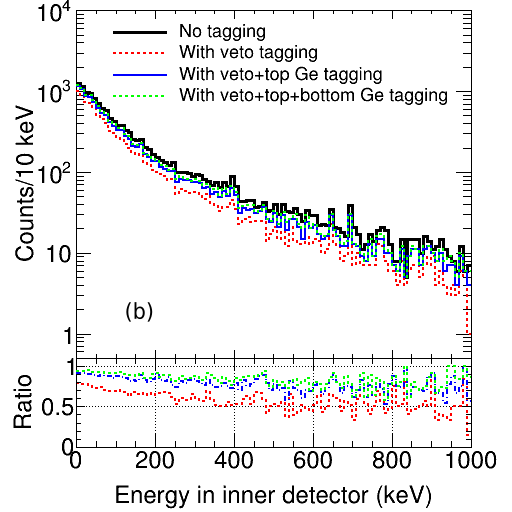}\\
\end{minipage}
\caption{Simulated energy deposition spectrum (upper panel) and the tagging efficiency (lower panel) as a function of energy deposition are shown in (a) germanium and (b) silicon inner detectors, respectively. Both the germanium and silicon detectors have a diameter of 25\,mm, and they are placed inside an annular germanium veto detector of 76 mm outer diameter and 28 mm inner diameter. The modules are then sandwiched between two germanium detectors of 76 mm diameter and 25 mm thick (see Fig. \ref{fig:Cs_simu}(c)). The black histogram represents the energy deposition distribution in the inner detector without considering any veto. The red dotted, blue solid and green dotted distribution represent energy deposition in the inner detector when tagged with the active veto, energy deposition when tagged with active veto and top germanium detector and energy deposition when tagged with the  veto, top and bottom germanium detectors respectively. Compton tagging efficiency with the active veto at the low energy ($<$ 100 keV) is $>$ 70\% which increases to $>$ 90\% after including top and bottom Ge detector as described in the figure.}
\label{fig:vetosim2}
\end{figure*}

An initial simulation was performed with a 25\,mm thick annular veto detector and a 4\,mm thick inner detector using GEANT4~\cite{geant4}. Fig. \ref{fig:Cs_simu}(a) shows a schematic diagram of the detector geometry. An isotropic point \textsuperscript{137}Cs source was simulated at a distance of 50 cm from the center of the detector along the detector plane to find the tagging efficiencies of Compton scattered photons, where tagging efficiency is defined as the fraction of events that deposit energy in the inner detector as well as in the veto. Fig. \ref{fig:Cs_simu}(b) shows the distribution of energy deposition (upper panel) and the tagging efficiency (lower panel) as a function of energy deposition in the inner detector. The figure shows that the tagging efficiency by the veto is expected to be between 50 - 80\%. A similar simulation was performed with the active veto detector sandwiched between two germanium detectors with a similar dimension using the expected background source for SuperCDMS SNOLAB with a threshold of 10 keV. Computer Aided Design or CAD software based representation of the detector stack shown in Fig. \ref{fig:Cs_simu}(c), is used in the GEANT4 simulation. The result is shown in Fig. \ref{fig:vetosim2}(a) which indicates that the tagging efficiency increases by placing the active veto detector with the germanium inner detector between two germanium detectors. The arrangement was sufficient to tag the Compton scatter events with an efficiency of approximately 90\% in the low energy ( $<$ 100 keV) region of interest. This provided confidence in the prototype design of the same thickness, hosting a smaller inner detector, as well as setting an energy threshold goal for the veto detector. These results were validated experimentally as shown in Section \ref{results}. The simulation was also performed with a silicon inner detector while the other detectors were made up of germanium. The result shown in Fig \ref{fig:vetosim2}(b) indicates that a germanium veto detector with a silicon inner detector is also capable of achieving similar tagging efficiency and would be very good for low mass dark matter searches. Fabrication of a veto detector with a silicon inner detector is being planned.

\section{Veto Detector Fabrication and Mounting}
\label{fab_testing}

The veto and inner detector pair prototype having mass $\sim$500 g and $\sim$10 g respectively, were produced and commissioned using the same methods as done for standard SuperCDMS detectors \cite{tamufab}. However, there were some engineering and design challenges to overcome in fabricating and mounting such a unique detector arrangement.

Three main challenges were addressed in the design of this setup:
\begin{enumerate}
    \item Fabricating an annular phonon-mediated veto detector with the ability to host an inner detector.
    \item Physically suspending the inner detector within the veto detector such that the inner detector is effectively cooled to operational temperatures.
    \item Connecting the inner detector circuit to existing cold electronics for signal readout.
\end{enumerate}

\subsection{Veto Detector Fabrication}
Photo-lithographically patterning the circuit on the detector surface is a challenge when fabricating an annular detector. The typical patterning process requires coating the surface with a layer of a photo-sensitive chemical called photo-resist. This is performed by rapidly spinning the substrate, relying on centrifugal motion to spread the chemical uniformly across the surface from the center. One concern was that this process would prove ineffective with a substrate that has a large hole in its center, resulting in non-uniform film thicknesses and therefore, non-uniform circuit patterns. In this paper, we have confirmed the possibility to pattern and test a detector fabricated in this way successfully.

The best method found to produce an annular detector was core-drilling (a process currently utilized by TAMU during detector fabrication \cite{tamufab}), proven not to degrade crystal characteristics for these purposes. During standard detector fabrication, detector substrates are delivered with a slightly larger diameter than their nominal dimension, then the core was drilled to the correct dimension with improved alignment to the crystal lattice. 
This is done before patterning the detector circuit. Using a pre-existing process such as core-drilling reduced R\&D work and risks, and is also able to neatly preserve the inner ``core'' which can be used for prototyping inner detectors as well as other test devices.

\subsection {Detector Circuit Design}
\begin{figure*}[ht!]
\begin{minipage}[b]{0.31\linewidth}
\centering
\includegraphics[width=\linewidth]{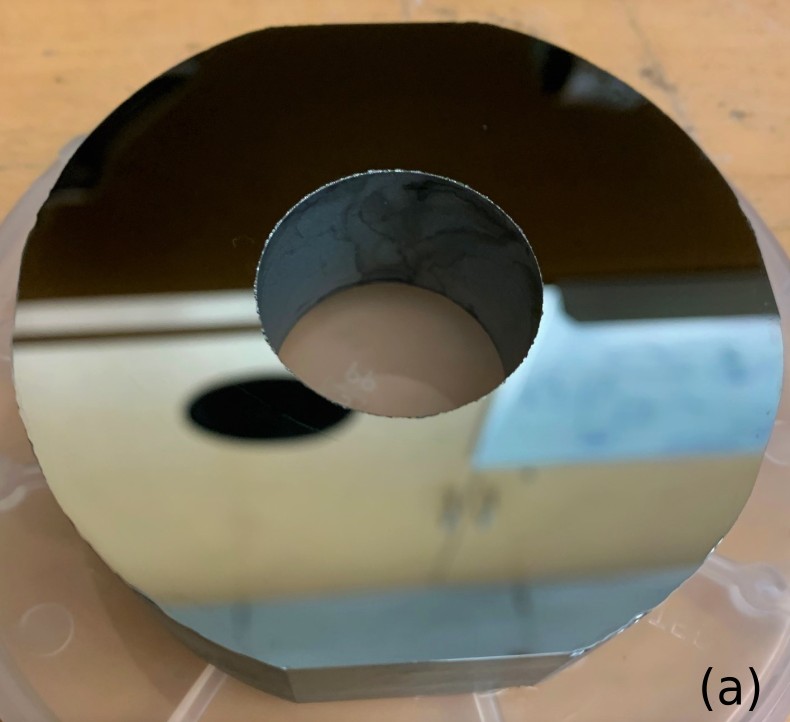}\\
\end{minipage}\hfill
\begin{minipage}[b]{0.31\linewidth}
\centering
\includegraphics[width=0.87\linewidth]{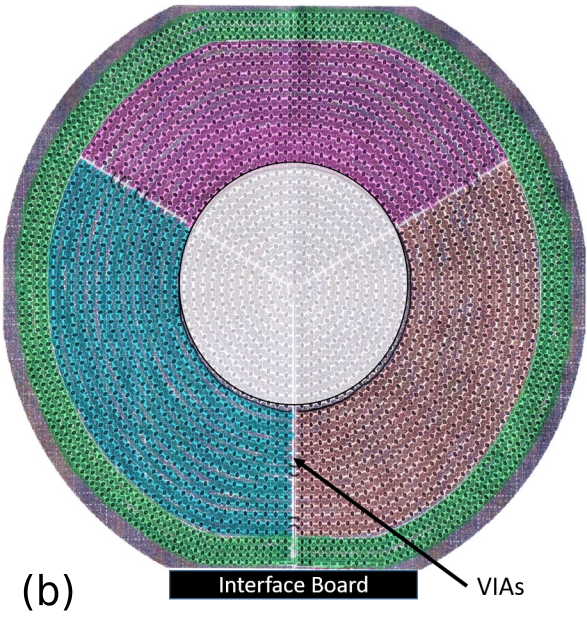}\\
\end{minipage}\hfill
\begin{minipage}[b]{0.31\linewidth}
\centering
\includegraphics[width=\linewidth]{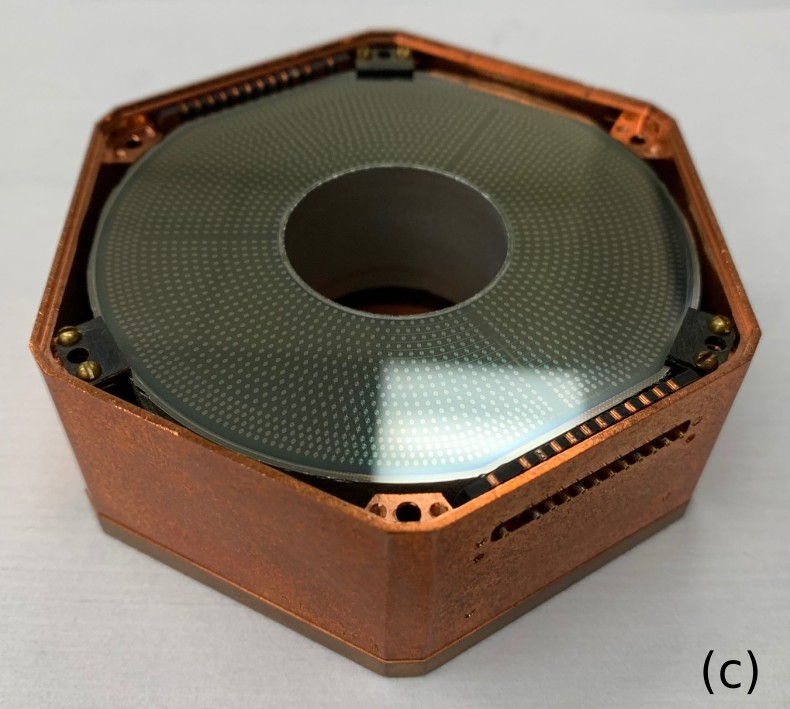}\\
\end{minipage}
\caption{(a) Veto detector crystal after polishing and core drilling, ready for patterning. (b) Graphic depiction of sensor layout showing the inner portion of sensors that is lost in the core drill region. Also shown are the aluminum lines (vias) to allow wiring of the inner detector using standard interface boards. The ``outer'' channel is shown in green, whereas the remaining sensors of the ``inner'' channels (blue, pink, and orange) were wired together in parallel to create one large inner channel. (c) The mounted veto detector, is ready for wire bonding followed by cryogenic testing.}
\label{fig:donutfab}
\end{figure*}

For this first prototyping work to demonstrate an inner active cryogenic veto, a pre-existing SuperCDMS mask design was used for the veto detector. Fig. \ref{fig:donutfab}(a) shows the veto detector crystal before patterning. The mask consists of an outer ring of sensors (the ``outer channel'') surrounding 3 inner sensors (see Fig. \ref{fig:donutfab}(b)). One of these inner sensors is on the opposite side of the crystal from the interface board where the detectors are wired to external electronics. Because of this, there are two aluminum lines on the mask/circuit that extend from the interface board to the center of the detector. These lines allowed the instrumentation of the inner detector as mentioned in Section \ref{mount}. Fig. \ref{fig:donutfab}(c) shows the veto detector after patterning. The circuit design for the inner detector was based on a different SCDMS mask that had 6 channels, including a circular inner channel. The mask was cropped to include only the inner channel which was then modified further to fit the constraints of the inner detector.
These circuit designs were chosen due to availability, time, and cost considerations. While these masks were originally designed for standard detector dimensions and wiring configurations, the alterations performed in this work deviate them further from optimal. Removing sensors and wiring channels in parallel affect the electrical characteristics of the channel, potentially reducing their performance significantly. For this reason, future work includes designing custom masks specifically for these geometries, providing a clear avenue for improvement.

\subsection{Mounting the Inner Detector}
\label{mount}

\begin{figure*}[ht!]
\centering
\begin{minipage}[b]{0.24\linewidth}
\centering
\includegraphics[width=\linewidth]{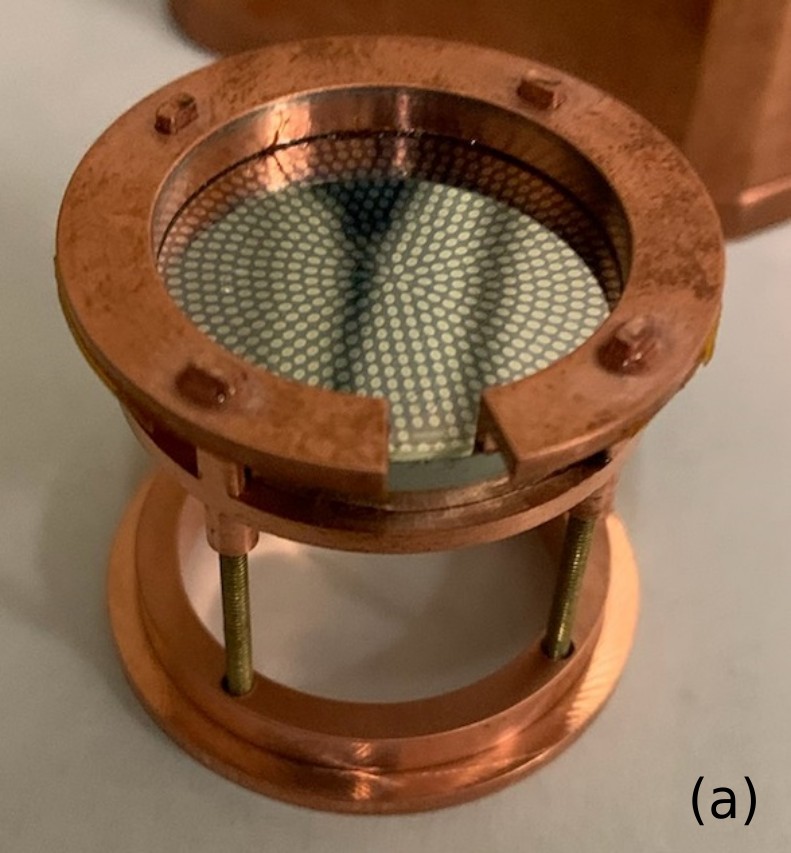}\\
\end{minipage}\hfill
\begin{minipage}[b]{0.35\linewidth}
\centering
\includegraphics[width=\linewidth]{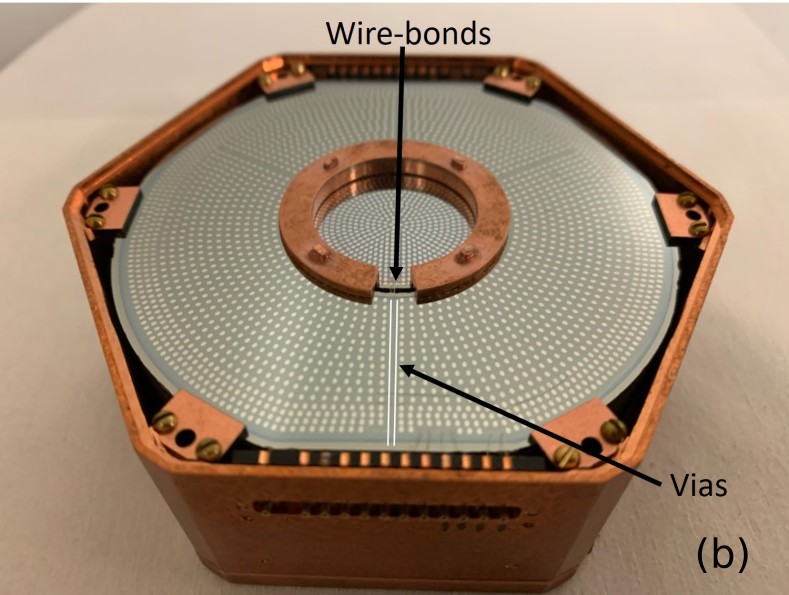}\\
\end{minipage}\hfill
\begin{minipage}[b]{0.35\linewidth}
\centering
\includegraphics[width=\linewidth]{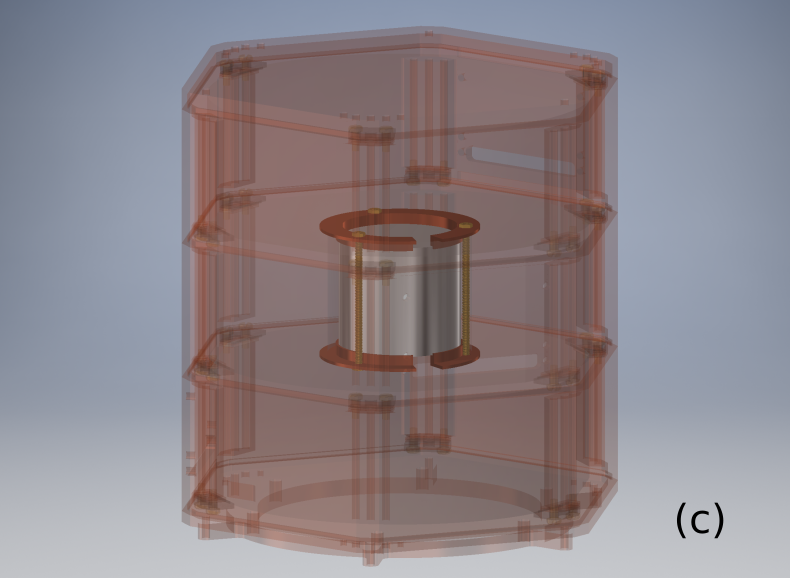}\\
\end{minipage}
\caption{Prototype detector mounting scheme. (a) Inner detector mounted in its copper nest. (b) Central detector nest mounted into veto detector, allowing the two detectors to be handled as a single typical 76 mm detector. (c) Proposed detector stack with 25 mm diameter by 22 mm thick inner detector with 4$\pi$ veto coverage.}
\label{fig:mount}
\end{figure*}

Another difficulty in implementing an internal veto detector is that it physically separates the inner detector from the housing. While this is beneficial, since the detector is then shielded from radiation emitted from the housing materials, it creates a challenge in regards to mounting the detector. The design chosen can be seen in Fig. \ref{fig:mount}(a) and (b). There are multiple motivations for this design: 
\begin{enumerate}
    \item Line of sight from the inner to veto detector.
    \item Line of sight from the inner detector to adjacent detectors above and below.
    \item Accommodations for wire bonding inner detector to veto.
    \item Physically combining the two detectors into one ``module''.
\end{enumerate}

The line of sight motivations stem purely from the coincident tagging efficiency. It is desirable to minimize any ``dead mass'' in the immediate vicinity of the detector. This prototype design coverage can be reduced as needed in later stages to improve that functionality. However, for prototype purposes, this design was sufficient for demonstration and a preliminary testing. The accommodations for wire bonding need to be considered in order to utilize aluminum vias patterned on the veto detector to allow signals to be read out from the inner detector. In this way, the inner detector is wire-bonded directly to the veto detector which is wire-bonded to the interface board on the housing. The gap in the mounting clamp ring allows for un-impeded wire-bonding from one crystal to the other. Combining the two detectors physically into one ``module'' allows one to treat the detector pair as if it were a single standard detector. From mounting to inspecting to wire-bonding and even circuit repair, the ``module'' can be treated as any other detector after the fabrication process.

A ``nest" design was chosen to satisfy these design constraints. It was constructed of an upper copper two-piece ``cage'' that holds the inner detector. The vertical ribs that connect the two pieces were initially angled slightly inward such that they press on the walls of the inner detector when it is installed. The two pieces were then epoxied together at their contact points on the top ring (see Fig. \ref{fig:mount}). This provided the thermal contact between the nest and the inner detector. To complete the thermal path to the veto detector, a lower ring was used to sandwich the veto detector and was tightened using brass threaded rods. This design allowed strong thermal contact throughout the cooling process and proved sufficient for prototyping purposes.

\section{Results}
\label{results}

With a successfully fabricated and mounted veto and inner detector module, various tests were performed to evaluate the performance of the prototype device. Of critical importance were the following:
\begin{itemize}
    \item Sufficiently cooling both detectors to operational temperatures.
    \item To achieve low baseline energy resolution for veto ($\sim$1 keV) and inner detector ($\sim$ 100 eV).
    \item Background reduction through coincident event tagging using the veto to limit inner detector data to ``single scatters''.
\end{itemize}
\subsection{Thermal Performance}
\label{thermal}

\begin{figure}
  \centering
    \includegraphics[width=0.8\linewidth]{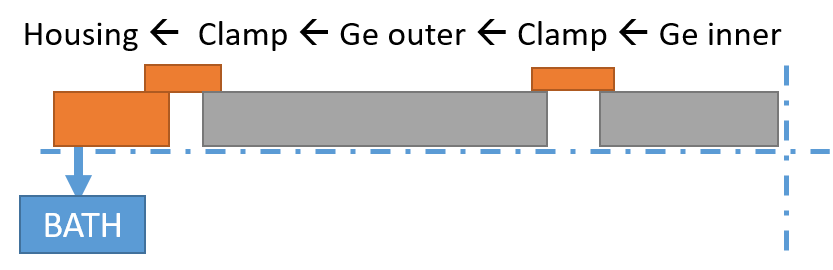}\\
\caption{Diagram showing the thermal conduction path from the inner detector to the refrigerator (bath). Dash-dotted lines indicate planes of symmetry. Orange material is copper, while gray is germanium.}
\label{fig:therm_path}
\end{figure}

\begin{figure*}[h!]
\centering
\begin{minipage}[b]{0.32\linewidth}
  \centering
\includegraphics[width=\linewidth]{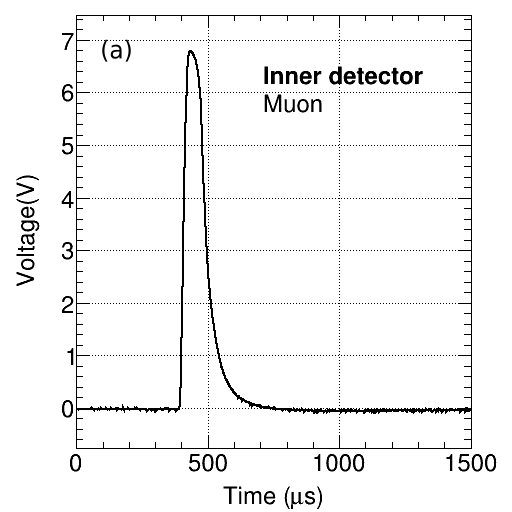}\\
\end{minipage}
\begin{minipage}[b]{0.32\linewidth}
  \centering
  \includegraphics[width=\linewidth]{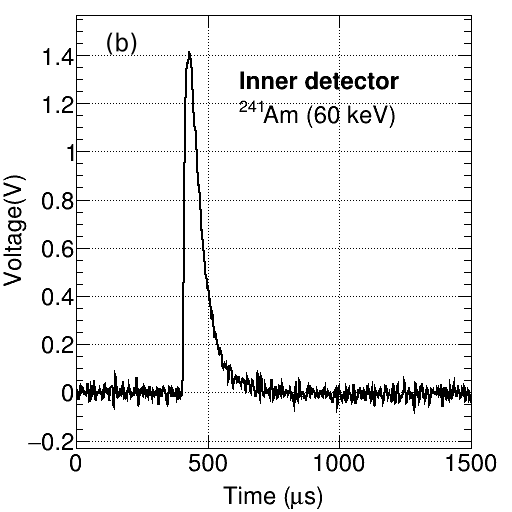}\\
\end{minipage}
\begin{minipage}[b]{0.32\linewidth}
\centering
\includegraphics[width=\linewidth]{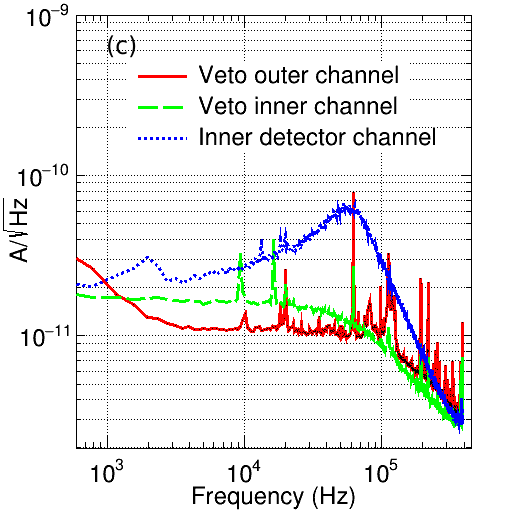}\\
\end{minipage}\\
\begin{minipage}{\linewidth}
  \centering
\begin{minipage}{0.32\linewidth}
  \centering
  \includegraphics[width=\linewidth]{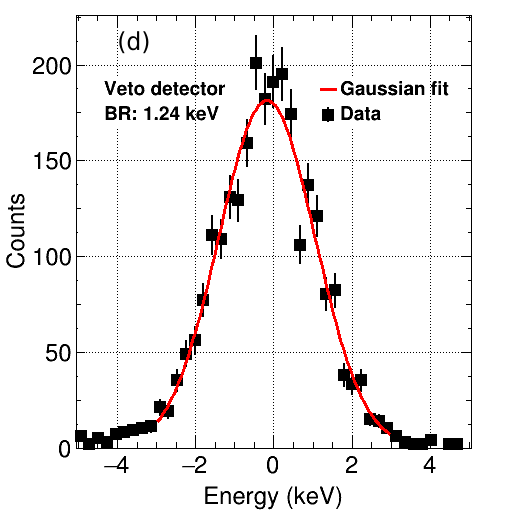}\\
\end{minipage}
\begin{minipage}{0.32\linewidth}
\centering
\includegraphics[width=\linewidth]{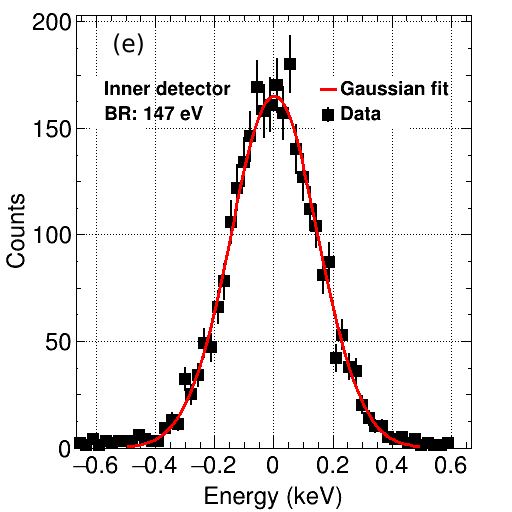}\\
\end{minipage}
\begin{minipage}{0.32\linewidth}
  \centering
  \caption{(a) Example muon pulse in the inner target detector is plotted as voltage vs time, showing rapid thermal recovery. The gain was set to minimum to avoid digitizer saturation from the muon pulses. (b) Example 60 keV pulse from the inner detector. (c) The TES current/$\sqrt{Hz}$ as a function of frequency is plotted which is known as Noise power spectral density (PSD). The PSD of inner and outer channels of the active veto detector and the inner detector channel. Noise PSD for inner detector channel behaves differently from veto detector channels as the relative bias applied on the Quasiparticle-trap-assisted Electrothermal-feedback Transition-edge-sensors (QET) electronics was higher. (d) Baseline resolution in units of keV is shown for veto detector. It shows baseline resolution of 1.24 keV while instrumented with the inner detector. (e) Baseline resolution of 147 eV measured by the inner detector.}
  \label{fig:new_results}
\end{minipage}
\end{minipage}

\end{figure*}

One concern mentioned previously is the ability to cool the inner detector and maintain an operational temperature ($\sim$tens of mK). The addition of the outer veto between the inner detector and its source of cooling power complicates this situation (see Fig. \ref{fig:therm_path}). The design adds multiple thermal impedances as well as possible avenues for coupling vibration.
As a qualitative study of the thermalization of the inner detector, the thermal recovery times from high energy depositions in the inner detector have been studied. Cosmic muons were used for this study, with approximately 3.3 MeV expected to be deposited by a muon in the inner detector. One would expect a relatively long thermal recovery time (compared to the typical pulse-width) if thermal sinking is poor, causing the detector temperature to be higher, and thus closer to the superconducting transition temperature of the sensors. An example of a muon pulse in the inner target detector is shown in Fig. \ref{fig:new_results}(a).  In this data, we found that the rise time for the muon pulse is $\sim$20 $\mu$s whereas the fall time is $\sim$90 $\mu$s indicating a very fast recovery time. This implies sufficient thermalization of the inner detector has been attained.

\subsection{Baseline Energy Resolution}
\label{BLR}

To understand the signal to noise (S/N) performance of the veto detector, the baseline resolution was measured. Electronic noise is one of the major sources of noise in the detector, causing a poor S/N ratio. Noise data was taken with random triggers to study the noise performance of the detectors. To calibrate the energy scale, an external \textsuperscript{57}Co and an internal \textsuperscript{241}Am source were used to illuminate the veto detector with known 122 keV and 60 keV gammas respectively. A pulse template fitting algorithm called ``Optimal filter (OF)'' \cite{OFmethod} was used to extract energy from raw pulses. An example of 60 keV pulse due to the \textsuperscript{241}Am source in the inner detector is shown in Fig. \ref{fig:new_results}(b). A pulse template was made by averaging over all such good pulses. From the noise data, we selected a few noise traces and obtained its power spectral density (PSD) (shown in Fig. \ref{fig:new_results}(c). The noise PSD, the pulse template and the raw data, all act as an input to the OF method which fits the template to the pulse and determines the pulse amplitude from the best fit result. The OF amplitude distribution of the entire data set can then be calibrated to the source using the known gamma energies. Baseline resolution is defined as the 1$\sigma$ width of the calibrated noise energy distribution which is calculated by fitting a gaussian to the data. The results can be seen in Fig. \ref{fig:new_results}(d). A baseline resolution of 1.24 keV was measured in the veto detector, while the inner detector mounted inside showed 147 eV baseline resolution (see Fig. \ref{fig:new_results}(e). Due to having low mass, the inner detector showed better S/N performance than the veto detector. A conservative energy threshold of the veto detector can be calculated as 5$\sigma$, which equates to $\sim$6 keV, well below the 10 keV energy threshold found sufficient in simulations (see Section \ref{simulation}).

The simulation and data analysis are performed for both the detectors operated at 0 V. It measures true recoil energy deposited by the particle in the detectors. Operating the detectors at a higher voltage will amplify the phonon signals linearly with the applied voltage through Neganov-Trofimov-Luke (NTL) \cite{NTL, Luke} effect. This will also improve the S/N ratio providing scope to explore lower energy regions (in eV scale) in future work.

\begin{figure*}[h!]
\centering
\begin{minipage}[b]{0.32\linewidth}
\centering
\includegraphics[width=\linewidth]{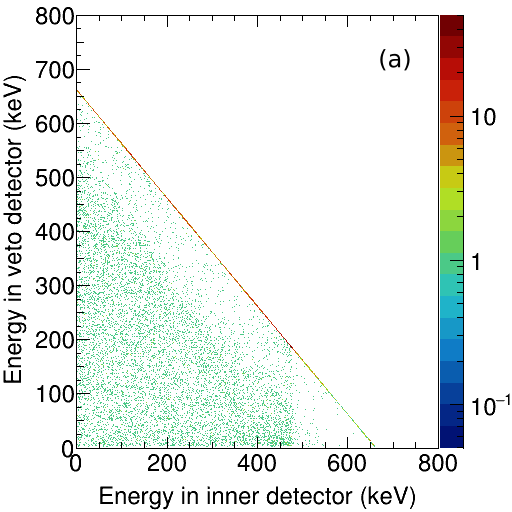}\\
\end{minipage}
\begin{minipage}[b]{0.32\linewidth}
\centering
\includegraphics[width=\linewidth]{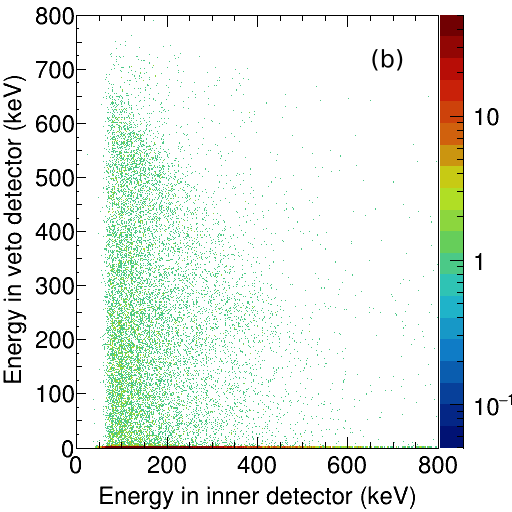}\\
\end{minipage}
\begin{minipage}[b]{0.32\linewidth}
\centering
\includegraphics[width=\linewidth]{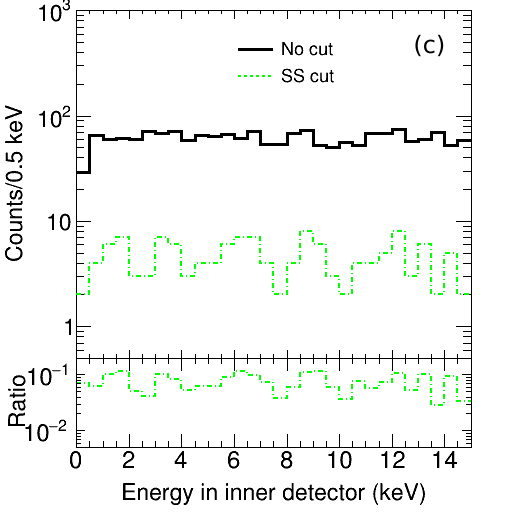}\\
\end{minipage}\\
\begin{minipage}{\linewidth}
  \centering
  \begin{minipage}{0.64\linewidth}
    \centering
    \begin{minipage}{0.8\linewidth}
\caption{(a) Simulation of events from a $^{137}$Cs source placed outside of the fridge, aimed at the side of the veto detector. The vertical axis represents energy deposition in veto detector and the horizontal axis reprsents the energy deposition in inner detector. (b) Experimental data showing coincident events due to Compton scatters from the same source at the same location. For this data the trigger threshold of the inner detector was raised to $\sim$70 keV to exclude events from the internal $^{241}$Am source. (c) Simulated background spectra (upper panel) in the inner detector using expected background source for SuperCDMS SNOLAB and (d) Background spectra measured by the inner detector (upper panel) at MINER experimental site with and without single scatter cut. In (c) and (d) the black histogram represents total energy deposition measured in the inner detector whereas the green histogram represents energy deposition by the single scatter events in inner detector}
\label{fig:cs137}
\end{minipage}
\end{minipage}
\begin{minipage}{0.32\linewidth}
\centering
\includegraphics[width=\linewidth]{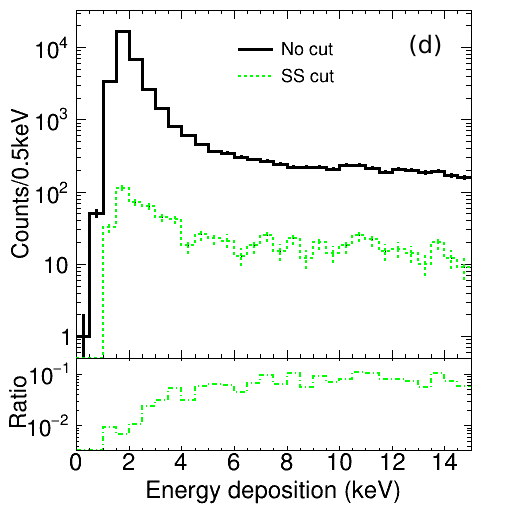}\\
\end{minipage}
\end{minipage}
\end{figure*}

\subsection{Background Reduction with the Veto Technique}

Further in-situ characterization of the active veto detector was performed at the MINER (Mitchell Institute Neutrino Experiment at Reactor \cite{minerbkg}) experimental site. The detector tower configuration inside the dilution fridge modeled using CAD is shown in Fig. \ref{fig:Cs_simu}(c) consisting of the inner and veto detector pair sandwiched between two 76 mm diameter and 25 mm thick germanium detectors. In order to confirm the ability to tag and veto Compton scatter events, a $^{137}$Cs source was collimated to shine 662 keV gammas on the side of the veto assembly. The setup was also simulated using GEANT4 to compare with the experimental results. The simulated and the experimental result are shown in Fig \ref{fig:cs137}(a) and (b) respectively. In the simulation detector resolution was not considered. For this reason, a sharp 662 keV gamma line was seen to be shared between inner and veto detector while in data we get a smeared distribution as seen in Fig \ref {fig:cs137}(b) that is qualitatively consistent. In the experiment, the threshold was raised to ignore 60 keV events from the internal $^{241}$Am calibration source. Background events are multiple scattering events and rare signals like CE$\nu$NS or dark matter interactions are basically single scatter (SS) events. Events with energy deposition above the threshold in the inner detector and energy deposition consistent with noise in the other detectors are considered as single scatter events (SS) in the analysis.
The detector setup was also simulated with the SuperCDMS SNOLAB background. It was seen that the background was reduced by an order of magnitude as shown in Fig. \ref{fig:cs137}(c). The experimental measurements are done at the MINER site, seen in Fig. \ref{fig:cs137}(d) also shows an order of magnitude reduction in the background.

\section{Conclusions}
\label{conclusions}
In this paper, we report the characterization of a newly developed phonon mediated veto detector prototype made up of germanium which shows an excellent reduction in ambient radiogenic background rate (dominantly gammas). We present here the detector fabrication process and detector mounting methodology addressing all the challenges in the design and setup. Sufficient thermalization of the detector is confirmed by studying the muon data. The veto detector shows a baseline resolution of 1.24 keV providing an energy threshold of $\sim$6 keV. We also show that the simulated response of the veto detector matches well with the experimental result. Furthermore, the detector is capable of rejecting three major backgrounds (surface betas, \textsuperscript{210}Pb, Compton scatters) which are expected to limit sensitivity for the SuperCDMS SNOLAB experiment. With the background rejection efficiency as high as 90$\%$  the active veto detector would be an excellent candidate for dark matter search and CE$\nu$NS experiments.
\section*{Acknowledgement} 
This work was supported by the DOE Grant Nos DE-SC0020097, DE-SC0018981, DE-SC0017859, and DE-SC0021051. We acknowledge the seed funding provided by the Mitchell Institute for the early conceptual and prototype development. We would like to acknowledge the support of DAE-India through the project Research in Basic Sciences - Dark Matter and SERB-DST-India through the J. C. Bose Fellowship. We also thank SuperCDMS SNOLAB Collaboration for the information on the projected sensitivity for the direct detection dark matter experiment.


\bibliography{mybibfile}

\end{document}